\documentclass[showpacs,preprintnumbers,amsmath,amssymb, twocolumn]{revtex4}
\usepackage{graphicx}
\usepackage{dcolumn}
\usepackage{bm}

\begin{document}

\title{Symmetry  Supporting a Transition to Zero Cosmological Constant State}

\author
{E. I. Guendelman \thanks{guendel@bgu.ac.il} and A.  B. Kaganovich
\thanks{alexk@bgu.ac.il}}
\address{Physics Department, Ben Gurion University of the Negev, Beer
Sheva 84105, Israel}

\date{\today}

\begin{abstract}
In a number of previous publications we demonstrated that the Two
Measures Field Theory (TMT) enables to resolve the old
cosmological constant (CC) problem avoiding the Weinberg's no-go
CC theorem and together with this TMT agrees with all tests of the
Einstein's general relativity and allows inflationary scenarios.
Analysis performed in the present paper shows that there exists an
intrinsic  symmetry of TMT which emerges in the $\Lambda =0$
ground state. This symmetry contains a subgroup of inversion of
the sign of the metric $g_{\mu\nu}\to -g_{\mu\nu}$ studied
recently by a number of authors as the symmetry imposing zero CC.
We show that realization of this idea in TMT is free of fine
tuning and has no problems typical to other approaches.

\end{abstract}

\maketitle

If the cosmological constant (CC) indeed equals zero the genuine
reason of the exact cancelation of all contributions to the vacuum
energy density is the most  intriguing puzzle of
 field theory and cosmology. One could expect that there is
going to be a chance to solve the old CC problem\cite{Weinberg1}
if the CC term is forbidden by some symmetry.  Recently two
related approaches have been discussed in literature: one is based
on the idea\cite{'t Hooft} to postulate a symmetry with respect to
a transformation towards imaginary values of the space-time
coordinates $x^{\mu}\to ix^{\mu}$;
other\cite{reflection1},\cite{reflection2} postulates  a symmetry
 transformation changing the sign of the metric tensor $g_{\mu\nu}\to
-g_{\mu\nu}$. These approaches encounter a number of very serious
problems concerning boundary conditions, surviving of  Standard
Model, etc..

A softer realization of the idea of a symmetry imposing zero CC
consists of the hypothesis\cite{'t Hooft} that such a symmetry
emerges only in the vacuum. However in this case the theory must
also dynamically provide a zero vacuum energy state without fine
tuning. The aim of the present paper is to demonstrate that the
Two Measures Field Theory (TMT) possesses exactly these features.
In fact, in a number of our papers\cite{GK1},\cite{GK2} has been
demonstrated that the TMT enables to resolve the old CC problem
avoiding the Weinberg's no-go CC theorem\cite{Weinberg1} and
together with this TMT agrees\cite{GK3} with all tests of the
Einstein's general relativity (GR) and allows inflationary
scenarios\cite{GK2}. In the present paper we show that the fine
tuning free transition to the $\Lambda =0$ ground state in TMT and
the emergence of an intrinsic TMT symmetry group (we call Einstein
symmetry) are effects mutually supporting each other.
Transformations of the Einstein symmetry involve the inversion
$g_{\mu\nu}\to -g_{\mu\nu}$. Here $g_{\mu\nu}$ is the metric
tensor used in the underlying action where the space-time turns
out to be non-Riemannian generically. During the process of the
transition to the $\Lambda =0$ vacuum state, the Einstein symmetry
manifests itself dynamically and in particular $g_{\mu\nu}$
oscillates about zero. But the metric tensor in the Einstein frame
is always regular.

TMT is a generally coordinate invariant theory where  the
effective action for 'gravity $+$ matter' at energies below the
Planck scale has  the form\cite{GK1}-\cite{GK3}
\begin{equation}
    S = \int L_{1}\Phi d^{4}x +\int L_{2}\sqrt{-g}d^{4}x
\label{S}
\end{equation}
 including two Lagrangians $ L_{1}$ and $L_{2}$ and two
volume measures $\sqrt{-g}$ and $\Phi$. One is the usual measure
of integration $\sqrt{-g}$ in the 4-dimensional space-time
manifold equipped with the metric $g_{\mu\nu}$. Another  is the
new measure of integration $\Phi$ in the same 4-dimensional
space-time manifold defined
 by means of  four scalar fields $\varphi_{a}$
\begin{equation}
\Phi
=\varepsilon^{\mu\nu\alpha\beta}\varepsilon_{abcd}\partial_{\mu}\varphi_{a}
\partial_{\nu}\varphi_{b}\partial_{\alpha}\varphi_{c}
\partial_{\beta}\varphi_{d}, \quad a=1,2,3,4
\label{Phi}
\end{equation}

One should pay attention  that {\it the measure} $\Phi$ {\it is
sign indefinite}. This fact turns out very important for problems
studied in the present paper  and it is a basic difference from
the attempts in Refs.\cite{reflection1},\cite{reflection2} to
realize a symmetry changing the sign of the volume element. In
Measure Theory\cite{signed} a sign indefinite measure is known as
"signed measure".

In  TMT it is assumed that the Lagrangian densities $ L_{1}$ and
$L_{2}$ are functions of all matter fields,  the metric, the
connection but not of the "measure fields" $\varphi_{a}$. In such
a case, i.e. when the measure fields  enter in the theory only via
the volume measure $\Phi$ in the form as in
  the action (\ref{S}),  the theory possesses
an infinite dimensional symmetry\cite{GK1}.  One can hope that
this symmetry should prevent emergence of a measure fields
dependence in $ L_{1}$ and $L_{2}$ after quantum effects are taken
into account.

Finally, TMT is formulated in the first order formalism, that is
all fields, including also metric, connection and the  measure
fields $\varphi_{a}$ are taken as independent dynamical variables.
All the relations between them are results of equations of motion.
In particular, the independence of the metric and the connection
means that the relation between them is not necessarily according
to Riemannian geometry. If we were proceed in the second order
formalism then the theory would be different from TMT we have
studied in Refs. \cite{GK1}-\cite{GK3}, see\cite{Comelli} and
references therein.

Except for the listed assumptions we do not make any changes as
compared with principles of the standard field theory in curved
space-time. In other words, all the freedom in constructing
different models in the framework of TMT consists of the choice of
the concrete matter content and the Lagrangians $ L_{1}$ and
$L_{2}$ that is quite similar to the standard field theory.

Since $\Phi$ is a total derivative, a shift of $L_{1}$ by a
constant, $L_{1}\rightarrow L_{1}+const$, has no effect on the
equations of motion. Similar shift of $L_{2}$ would lead to the
change of the constant part of the Lagrangian coupled to the
volume element $\sqrt{-g}d^{4}x $. In the standard GR, this
constant term is the CC. However working with the metric
$g_{\mu\nu}$ of the underlying action (\ref{S}) we deal with a
non-Riemannianin space-time. This is why the relation between the
constant
 term of $L_{2}$ and the physical CC is very non
trivial.

Varying the measure fields $\varphi_{a}$, we get
$B^{\mu}_{a}\partial_{\mu}L_{1}=0$ where
$B^{\mu}_{a}=\varepsilon^{\mu\nu\alpha\beta}\varepsilon_{abcd}
\partial_{\nu}\varphi_{b}\partial_{\alpha}\varphi_{c}
\partial_{\beta}\varphi_{d}$.
Since $Det (B^{\mu}_{a}) = \frac{4^{-4}}{4!}\Phi^{3}$ it follows
that if $\Phi\neq 0$ (i.e. except for possible topological defects
where $\Phi$ might be zero)
\begin{equation}
 L_{1}=sM^{4} =const
\label{varphi}
\end{equation}
where $s=\pm 1$ and $M$ is a constant of integration with the
dimension of mass. Variation of the metric $g^{\mu\nu}$ gives
\begin{equation}
\xi\frac{\partial L_1}{\partial g^{\mu\nu}}+\frac{\partial
L_2}{\partial g^{\mu\nu}}-\frac{1}{2}g_{\mu\nu}L_2 =0, \quad
\text{where}\quad \xi\equiv \frac{\Phi}{\sqrt{-g}}
\label{g-mu-nu-varying}
\end{equation}
is the scalar field build of the scalar densities $\Phi$ and
$\sqrt{-g}$.

We study  models with  the Lagrangians of the form
\begin{equation}
L_1=-\frac{1}{\kappa}R(\Gamma, g)+L_1^m, \quad
L_2=-\frac{b_g}{\kappa}R(\Gamma, g)+L_2^m \label{L1L2}
\end{equation}
 where $\Gamma$ stands
for affine connection, $R(\Gamma,
g)=g^{\mu\nu}R_{\mu\nu}(\Gamma)$,
$R_{\mu\nu}(\Gamma)=R^{\lambda}_{\mu\nu\lambda}(\Gamma)$ and
$R^{\lambda}_{\mu\nu\sigma}(\Gamma)\equiv \Gamma^{\lambda}_{\mu\nu
,\sigma}+ \Gamma^{\lambda}_{\alpha\sigma}\Gamma^{\alpha}_{\mu\nu}-
(\nu\leftrightarrow\sigma)$. Dimensionless factor $b_g$ in front
of $R(\Gamma, g)$ in $L_2$ appears because there is no reason for
couplings  of the scalar curvature to the measures $\Phi$ and
$\sqrt{-g}$ to be equal. We choose $b_g>0$ and $\kappa =16\pi G$.
$L_1^m$ and $L_2^m$ are the matter Lagrangians which can contain
all possible terms compatible with postulated symmetries.

Since the measure $\Phi$ is sign indefinite, the total volume
measure $(\Phi +b_g\sqrt{-g})$ in the gravitational term
$-\kappa^{-1}\int R(\Gamma, g)(\Phi +b_g\sqrt{-g})d^4x$ in the
action (\ref{S}) is generically also sign indefinite.

Applying the Palatini formalism  one can show\cite{GK1}-\cite{GK3}
that in addition to the Christoffel's connection coefficients, the
resulting relation between connection and metric includes also the
gradient of $\xi$. Consequently geometry of the space-time with
the metric $g_{\mu\nu}$ is non-Riemannian if $\xi\neq const.$. In
particular, the covariant derivative of $g_{\mu\nu}$ with this
connection is nonzero (nonmetricity). The gravity and matter field
equations obtained by means of the first order formalism contain
both $\xi$ and its gradient as well. It turns out that at least at
the classical level, the measure fields $\varphi_a$ affect the
theory only through the scalar field $\xi$.

For the class of models (\ref{L1L2}), the consistency of
Eqs.(\ref{varphi}) and (\ref{g-mu-nu-varying}) implies the
following algebraic equation
\begin{equation}
(\xi -b_g)(sM^4-L_1^m)+g^{\mu\nu}\left(\xi \frac{\partial
L_{1m}}{\partial g^{\mu\nu}}+\frac{\partial L_2^m}{\partial
g^{\mu\nu}}\right)-2L_2^m=0 \label{Constr-original}
\end{equation}
which determines  $\xi(x)$ as a local function of matter fields.
Note that the geometrical object $\xi(x)$ has no its own dynamical
equation of motion and its space-time behavior is totally
determined by the matter fields dynamics via
Eq.(\ref{Constr-original}). Together with this, since  $\xi$
enters into the matter field equations, it generically has
straightforward effects on the matter dynamics through the forms
of potentials, variable fermion masses and
selfinteractions\cite{GK1}-\cite{GK3}.

With the new metric defined by the transformation
\begin{equation}
\tilde{g}_{\mu\nu}=(\xi +b_{g})g_{\mu\nu}, \label{gmunuEin}
\end{equation}
the connection $\Gamma^{\lambda}_{\mu\nu}$ becomes equal to the
Christoffel connection coefficients of the metric
$\tilde{g}_{\mu\nu}$ and the space-time turns into (pseudo)
Riemannian. This is why the set of dynamical variables using the
metric $\tilde{g}_{\mu\nu}$ we call the Einstein frame. It is very
important that {\it the transformation} (\ref{gmunuEin}) {\it is
not a conformal} one since $(\xi +b_{g})$ is sign indefinite. But
$\tilde{g}_{\mu\nu}$ is the regular pseudo-Riemannian metric.
Gravitational equations (\ref{g-mu-nu-varying}) in the Einstein
frame take canonical GR form\cite{GK1}-\cite{GK3}
\begin{equation}
G_{\mu\nu}(\tilde{g}_{\alpha\beta})=\frac{\kappa}{2}T_{\mu\nu}^{eff}
\label{gef}
\end{equation}
where  $G_{\mu\nu}(\tilde{g}_{\alpha\beta})$ is the Einstein
tensor in the Riemannian space-time with the metric
$\tilde{g}_{\mu\nu}$ and the energy-momentum tensor
$T_{\mu\nu}^{eff}$  depends on $\xi$
  which is a local function of matter
  fields resulting from Eq.(\ref{Constr-original}) rewritten in the Einstein frame.

Let us now turn to intrinsic symmetry of TMT which can reveal
itself in a model with only the measure $\Phi$. Indeed, if in
Eq.(\ref{L1L2}) $b_g=0$ and $L_2^m\equiv 0$ then
Eq.(\ref{Constr-original}) reads
\begin{equation}
L_1^m-g^{\mu\nu}\frac{\partial L_1^m}{\partial g^{\mu\nu}}=sM^4,
\quad {\text if} \quad\xi\neq 0. \label{Constr-L2=0}
\end{equation}
If in addition $L_1^m$ is homogeneous of degree 1 in $g^{\mu\nu}$
then the integration constant $M$ must be zero. The simplest
example of a model for $L_1^m$ satisfying this property is the
massless scalar field. In such a case the theory is invariant
under the local transformations
 \begin{eqnarray}
&&\Phi(x)\rightarrow\Phi^{\prime}(x)=J(x)\Phi(x),\nonumber\\
&&g_{\mu\nu}(x)\rightarrow
g^{\prime}_{\mu\nu}(x)=J(x)g_{\mu\nu}(x) \label{LES}
\end{eqnarray}
where $J(x)=
Det(\frac{\partial\varphi^{\prime}_{a}}{\partial\varphi_{b}})$ is
the Jacobian of a transformation
$\varphi_{a}\rightarrow\varphi^{\prime}_{a}=
\varphi^{\prime}_{a}(\varphi_{b})$ in the space of the scalar
fields $\varphi_{a}$. This symmetry was studied in earlier
pulications\cite{GK1} where we called it the local Einstein
symmetry (LES).

Consider now linear transformations in the space of the scalar
fields $\varphi_{a}$
\begin{equation}
\varphi_{a}\rightarrow\varphi^{\prime}_{a}= A_a^b\varphi_{b}+C_b,
\quad a,b=1,2,3,4 \label{linear-trans}
\end{equation}
where $A_a^b=constants$, $C_b=constants$. Then LES (\ref{LES}) is
reduced to transformations of the global Einstein symmetry (GES)
with $J=det(A_a^b)=const$. The  feature of the Einstein symmetry
we did not discuss before consists in the existence of $\Bbb Z_2$
subgroup of the sign inversions when $J=-1$:
\begin{equation}
\Phi\rightarrow -\Phi, \quad g_{\mu\nu}\rightarrow
-g_{\mu\nu}\label{reflection}
\end{equation}
that, as we will see below, is very important for understanding
what happens in the transition to a $\Lambda =0$ state.

LES as well as GES appear to be explicitly broken if $L_1^m$ is
not a homogeneous function of degree 1 in $g^{\mu\nu}$, for
example as in the model where $L_1^m$ describes a scalar field
with a nontrivial potential. The Lagrangian $L_2^m$ generically
breaks the Einstein symmetry too. The transformation of GES
originated by the infinitesimal linear transformations
$\varphi_a(x)\rightarrow\varphi^{\prime}_a(x)=
(1+\epsilon/4)\varphi_a(x)$, $\epsilon =const.$, yields the
following variation of the action $(\ref{S})$ written in the form
$S=\int{\cal L}d^4x$ where ${\cal L}=\Phi L_1+\sqrt{-g}L_2$:
\begin{equation}
\delta S=\int\left[-\frac{\partial{\cal L}}{\partial
g^{\mu\nu}}g^{\mu\nu}+
L_1\frac{\partial\Phi}{\partial\varphi_{a,\mu}}\varphi_{a,\mu}\right]\epsilon
d^4x. \label{deltaS}
\end{equation}
The first term in (\ref{deltaS}) equals zero on the mass shell
giving the gravitational equation (\ref{g-mu-nu-varying}); recall
that we proceed in the first order formalism. Integrating the
second term by part, using Eq.(\ref{varphi})  and the definition
(\ref{Phi}) of the measure $\Phi$, we reduce the variation
(\ref{deltaS}) to $\delta S=\epsilon\int\partial_{\mu}j^{\mu}d^4x$
where $\partial_{\mu}j^{\mu}=sM^4\Phi$ and
$j^{\mu}=sM^4B^{\mu}_a\varphi_a$. In the presence of topological
defects with $\Phi =0$, Eq.(\ref{varphi}) does not hold anymore
all over space-time, and one should keep $L_1$ in the definition
of the current: $j^{\mu}=L_1B^{\mu}_a\varphi_a$.

To present the result in the generally coordinate invariant form
one has to use the covariant divergence. However when doing this
using the original metric we encounter the non-metricity problem.
It is much more transparent to use the Einstein frame
(\ref{gmunuEin}) where the space-time becomes pseudo-Riemannian
and the covariant derivative of the metric $\tilde{g}_{\mu\nu}$
equals zero identically. Thus with the definition
$j^{\mu}=\sqrt{-\tilde{g}}J^{\mu}$, using the definition of $\xi$
in Eq.(\ref{g-mu-nu-varying}) and the transformation to the
Einstein frame (\ref{gmunuEin}) we obtain
\begin{equation}
\tilde{\nabla}_{\mu}J^{\mu}\equiv\frac{1}{\sqrt{-\tilde{g}}}\partial_{\mu}
\left(\sqrt{-\tilde{g}}J^{\mu}\right)=sM^4\frac{\xi}{(\xi +b_g)^2}
\label{covar-conserv-J}
\end{equation}

As one should expect, when $L_2\equiv 0$ and $L_1^m$ is
homogeneous of degree 1 in $g^{\mu\nu}$, i.e. in the case of
unbroken GES, the current is conserved because in this case the
integration constant $M=0$. Let us consider now more realistic
model including gravity as in Eqs.(\ref{L1L2}) and a scalar field
$\phi$ with
\begin{equation}
L_1^m=\frac{1}{2}g^{\mu\nu}\phi_{,\mu}\phi_{,\nu}-V_1(\phi),\quad
L_2^{m}=\frac{b_{\phi}}{2}g^{\mu\nu}\phi_{,\mu}\phi_{,\nu}-V_2(\phi)
\label{model}
\end{equation}
The appearance of the dimensionless factor $b_{\phi}$ is explained
by the fact that normalizing all the fields such that their
couplings to the measure $\Phi$ have no additional factors, we are
not able in general to provide the same in the terms describing
the appropriate couplings to the measure $\sqrt{-g}$. Generically
$b_{\phi}\neq b_g$ that yields a nonlinear kinetic term (i.e. the
$k$-essence type dynamics) in the Einstein frame\cite{GK2}. But
for purposes of the present paper it is enough to take a
simplified model with $b_{\phi}= b_g$ (which is in fact a fine
tuning) since the nonlinear kinetic term has no qualitative effect
on the CC problem.

The TMT procedure consists of using the least action principle
with respect to  all the dynamical variables, as the first step,
and in the second step one should rewrite all equations of motion
in the Einstein frame.
The field equations of the
model (\ref{model}) written in the Einstein frame include the
Einstein equations (\ref{gef}) with the energy-momentum tensor
\begin{equation}
T_{\mu\nu}^{eff}=\phi_{,\mu}\phi_{,\nu}-\frac{1}{2}
\tilde{g}_{\mu\nu}\tilde{g}^{\alpha\beta}\phi_{,\alpha}\phi_{,\beta}
+\tilde{g}_{\mu\nu}V_{eff}(\phi;\zeta,M),
 \label{Tmunueff}
\end{equation}
where putting $M$ in the arguments of $V_{eff}$ we indicate
explicitly that $V_{eff}$ incorporates our choice for the
integration constant $sM^4$
\begin{equation}
 V_{eff}(\phi;\zeta ,M)=
\frac{b_{g}\left[sM^{4}+V_{1}\right] -V_{2}}{(\zeta +b_{g})^{2}},
\label{Veff1}
\end{equation}
 and the $\phi$-equation
\begin{equation}
\frac{1}{\sqrt{-\tilde{g}}}\partial_{\mu}\left(\sqrt{-\tilde{g}}
\tilde{g}^{\mu\nu}\partial_{\nu}\phi\right)+\frac{\partial
V_{eff}}{\partial\phi}=0.
 \label{phi-eq}
\end{equation}
  As we
will see soon, the $\xi$-dependence of $V_{eff}(\phi;\xi ,M)$ in
the form of inverse square like $(\xi +b_{g})^{-2}$ has a key role
in the resolution of the old CC problem in TMT. One can show that
if quantum corrections to the underlying action generate
nonminimal coupling like  $\propto R(\Gamma,g)\phi^2$ in both
$L_1$ and $L_2$, the general form of the $\xi$-dependence of
$V_{eff}$ remains similar: $V_{eff}\propto (\xi +f(\phi))^{-2}$,
where $f(\phi)$ is a function. The fact that only such type of
$\xi$-dependence emerges in $V_{eff}(\phi;\zeta ,M)$, and a
$\xi$-dependence is absent for example in the numerator of
$V_{eff}(\phi;\xi ,M)$, is a direct result of our basic assumption
that $L_1$ and $L_2$ are independent of the measure fields.

The scalar field $\xi$ in $V_{eff}(\phi;\zeta,M)$ is determined as
the solution of Eq.(\ref{Constr-original}) which reads now
\begin{equation}
(\xi -b_g)\left[sM^4+V_1(\phi)\right]+2V_2(\phi)=0
\label{xi-solution}
\end{equation}
and it results in the finite form of $V_{eff}(\phi)$:
\begin{equation}
 V_{eff}(\phi)=
\frac{(sM^{4}+V_{1})^2}{4[b_g(sM^4+V_1)-V_{2}]}. \label{Veff2}
\end{equation}

For an arbitrary nonconstant function $V_1(\phi)$ there exist
infinitely many values of the integration constant $sM^4$ such
that  $V_{eff}(\phi)$ has the {\it absolute minimum} at some
$\phi=\phi_0$ with $V_{eff}(\phi_0)=0$ (provided
$b_g[sM^4+V_1(\phi)]-V_2(\phi)>0$). This happens {\bf without any
sort of fine tuning} as $sM^4+V_1(\phi_0)=0$ . When the scalar
field $\phi$ rolls down its absolute minimum, the scalar
$|\xi|\to\infty$ and, according to Eq.(\ref{covar-conserv-J}),
$\tilde{\nabla}_{\mu}J^{\mu}\to 0$. Thus, the
 {\em GES explicitly broken in the underlying action, emerges in
the vacuum which, as it turns out, has zero energy density. And
vice versa, emergence of GES due to $|\xi|\to\infty$ implies,
according to Eq.(\ref{Veff1}), a transition to a $\Lambda =0$
state}.

Other way to understand what happens when the dynamical evolution
of the gravity$+$scalar field $\phi$ -system pushes
$|\xi|\equiv|\Phi|/\sqrt{-g}\to\infty$, is to look at the
underlying action (\ref{S}),(\ref{L1L2}),(\ref{model}). It is
evident that in this limit the second term in (\ref{S}) becomes
negligible in comparison with the first one, and therefore the
only remaining term which breaks the GES is $-\int V_1(\phi)\Phi
d^4x$. However, as $|\xi|\to\infty$, the consistency condition
(\ref{Constr-original}) takes the form (\ref{Constr-L2=0}) which
is equivalent to the condition $V_1(\phi_0)+sM^4=0$ where $\phi_0$
is such a value of $\phi$ that the integration constant $sM^4$
compensates $V_1(\phi_0)$ exactly. The constancy of $\phi$ means
that this is the stable vacuum state, and as we have seen above it
has zero energy density. On the other hand, the appearance of the
integration constant $sM^4$ (see Eq.(\ref{varphi})) is the
intrinsic feature of TMT resulting from the presence of the
measure $\Phi$ in the underlying action. Thus we conclude that
{\it the effect of the measure $\Phi$ in the transition to a
$\Lambda =0$ state consists in the dynamical damping of all terms
breaking the GES}.

To see in more details characteristic features of the transition
to a $\Lambda =0$ state it is worth to look at the appropriate
cosmological solution in a simple model. Let us consider a model
(\ref{L1L2}), (\ref{model}) with $V_1=\frac{1}{2}\mu_1^2\phi^2$,
$V_2=V_2^{(0)}+\frac{1}{2}\mu_2^2\phi^2$; notice that adding a
constant to $V_1$ does not effect equations of motion. We take
negative integration constant, i.e. $s=-1$, and the only
restriction on the values of the integration constant $M$ and the
parameters is that denominator in (\ref{Veff2}) is positive.

Consider spatially flat FRW universe with the metric in the
Einstein frame $\tilde{g}_{\mu\nu}=diag(1,-a^2,-a^2,-a^2)$, where
$a=a(t)$ is the scale factor. Without any sort of fine tuning, the
cosmological solution ends with the transition to a $\Lambda =0$
state. The characteristic features of this transition are
presented in Figs.1 and 2. In the phase plane of the scalar $\phi$
(Fig.2) one can see the oscillatory behavior of $\phi$ around the
absolute minimum $\phi_0$ where the energy density equals zero
(for illustrative purposes the parameters are chosen such that
$V_{eff}=(M^2/2b_g)(\phi^2-M^2)^2/(\phi^2+4M^2)$ and $\phi_0=\pm
M$). Four graphs in Fig.1 show behavior of components of the
metric $g_{\mu\nu}=(\xi +b_g)^{-1}\tilde{g}_{\mu\nu}$ (used in the
underlying action where the space-time is non-Rimannian), of the
measure $\Phi$ and the scalar in the r.h.s. of
Eq.(\ref{covar-conserv-J}).  All these quantities tend zero via
oscillations around zero. {\bf Solution with such type of
oscillations becomes possible because the GES and its subgroup of
the sign inversions} (\ref{reflection}) {\bf emerges near the
$\Lambda =0$ vacuum state}. Notice that  the measure $\Phi$ and
the metric $g_{\mu\nu}$ pass zero  only in a discrete set of
moments during the transition to the $\Lambda =0$ state. Therefore
there is no problem with the condition $\Phi\neq 0$ used in the
solution. Also there is no problem with singularity of
$g^{\mu\nu}$ in the underlying action  since $\Phi g^{\mu\nu}$ is
always finite. The metric in the Einstein frame
$\tilde{g}_{\mu\nu}$ is always regular because irregularity of
$g_{\mu\nu}$ is compensated in Eq.(\ref{gmunuEin}) by singularity
of the ratio of two measures $\xi\equiv\Phi/\sqrt{-g}$.

Qualitatively the same relation between the fine tuning free
transition to $\Lambda =0$ state and emergence of the GES holds in
all  TMT models of the form (\ref{S}),(\ref{L1L2}),(\ref{model})
with a non-trivial dynamics and, in particular, in the scale
invariant models\cite{G},\cite{GK2}. This will be shown in a
longer paper.

We acknowledge M. Lin, E. Nissimov and S. Pacheva for
clarification of the meaning of signed measure. One of us (E.G.)
thanks M. Duff for discussions on the subject of the paper.

\begin{center}
\begin{figure}[htb]
\includegraphics[width=16.0cm,height=12.0cm]{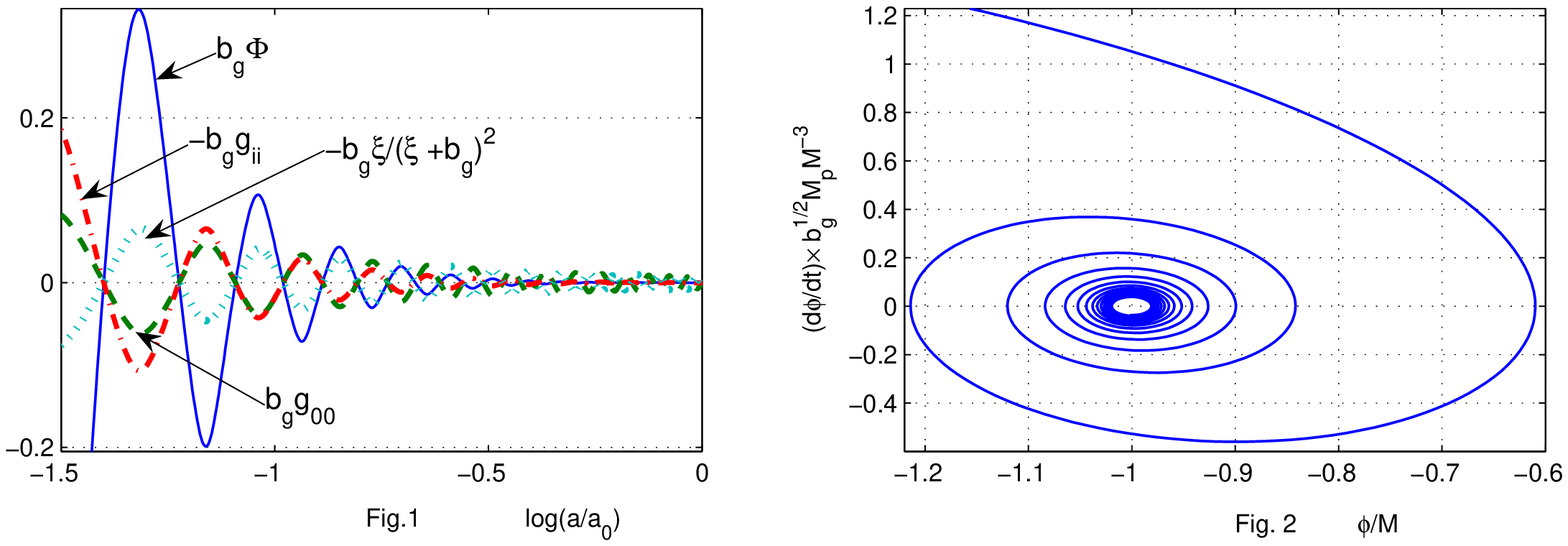}
\nonumber
\end{figure}
\end{center}



\begin{thebibliography}{99}

\bibitem{Weinberg1}
S. Weinberg, Rev. Mod. Phys. {\bf 61}, 1 (1989).

\bibitem{'t Hooft}
S. Nobbenhuis, Found. Phys. {\bf 36}, 613 (2006); G.'t Hooft, S.
Nobbenhuis, Class. Quant. Grav. {\bf 23}, 3819 (2006).

\bibitem{reflection1}
R. Erdem, Phys. Lett. B{\bf 621}, 11 (2005); ibid. B{\bf 639}, 348
(2006); J. Phys. A{\bf 40}, 6945 (2007).

\bibitem{reflection2} M.J. Duff, J. Kalkkinen, Nucl. Phys. B{\bf 758}, 161
(2006); B{\bf 760}, 64 (2007).

\bibitem{GK1} See E.I. Guendelman and A.B. Kaganovich, {\it Phys.
 Rev.} D{\bf 60}, 065004 (1999) and references therein.

\bibitem{GK2} See E.I. Guendelman and A.B. Kaganovich,  Phys. Rev. D{\bf }75, 083505
(2007) and references therein.

\bibitem{GK3}
E. Guendelman, A. Kaganovich,  Int. J. Mod. Phys. A{\bf 17}, 417
(2002); {\bf 21}, 4373 (2006); arXiv: 0704.1998 [gr-qc].

\bibitem{signed} D.L. Cohn, {\it Measure
Theory}, Birkhauser, Boston, 1993.

\bibitem{Comelli}
D. Comelli, arXiv:0704.1802 [gr-qc].

\bibitem{G}
E.I. Guendelman, Mod. Phys. Lett. A{\bf 14}, 1043 (1999).

\end{thebibliography}
\end{document}